# Orbital driven impurity spin effect on the magnetic order of quasi-three dimensional cupric oxide

**B G Ganga[1], P N Santhosh[1] and B R K Nanda[2*]**

[1] Low Temperature Physics Lab, Department of Physics, IIT Madras, Chennai, 600036 India

[2] Condensed Matter Theory and Computational Lab, Department of Physics, IIT Madras, Chennai, 600036 India

[*] Email: nandab@iitm.ac.in

**Abstract** Density functional calculations are performed to study the magnetic order of the severely distorted square planar cupric oxide (CuO) and local spin disorder in it in the presence of the transition metal impurities M (= Cr, Mn, Fe, Co and Ni). The distortion in the crystal structure, arisen to reduce the band energy by minimizing the covalent interaction, creates two crisscrossing zigzag spin-1/2 chains. From the spin dimer analysis we find that while the spin chain along $[10\bar{1}]$ has strong Heisenberg type antiferromagnetic coupling (J ~ 127 meV), along [101] it exhibits weak, but robust, ferromagnetic coupling (*J* ~ 9 meV) mediated by reminiscent *p-d* covalent interactions. The impurity effect on the magnetic ordering is independent of M and purely orbital driven. If the given spin-state of M is such that the $d_{x^2-y^2}$ orbital is spin-polarized, then the original long-range ordering is maintained. However, if $d_{x^2-y^2}$ orbital is unoccupied, the absence of corresponding covalent interaction breaks the weak ferromagnetic coupling and a spin-flip takes place at the impurity site leading to breakdown of the long range magnetic ordering.

## 1. Introduction

Most of the TMOs such as monoxides (MO), perovskites (AMO$_3$) and Ruddlesden-Popper series ($A_{n-1}A'_2M_nO_{3n+1}$), spinel compounds crystallize with M-O-M layers as well as MO symmetric (octahedral, tetrahedral etc.) complexes. Minor structural distortions in these systems may occur depending on the Jahn-Teller (JT) distortion modes to remove the band degeneracy. The perovskite compound LaMnO$_3$ is a classic example, where doubly degenerate itinerant $e_g$ bands split into two non-degenerate bands through breathing and stretching distortion of MnO$_6$ octahedra [1].

The cupric oxide, unlike the other monoxides, neither stabilizes with Cu-O-Cu planes nor forms CuO$_6$ octahedral complexes. It stabilizes in a monoclinic lattice with space group symmetry *C2/c* [2] as shown in Figure 1d. The structure consists of two criss-crossed zig-zag Cu chains extended along [101] and $[10\bar{1}]$. Each Cu forms a weakly distorted square planar CuO$_4$ complex. However, the neighboring CuO$_4$ complexes are tilted with each other and connected through the edge sharing oxygen. The displacement of Oxygen ions in these complexes lead to electric polarization of P ~ 0.02μC/cm$^2$ along the y-axis [3] and thereby makes the compound a promising multiferroic material [4-6]. In addition, the Cu-Cu non-planar bond length (along b-axis) is close to 1.77 Å whereas in other monoxides and perovskites two successive MO planes are separated by a distance of order of 4 Å [7, 8]. Therefore, even though it is not a planar structure, it cannot be considered as a perfectly three dimensional compound as assumed by many [9-11]. Hence, CuO provides a proper platform to study the electronic and magnetic structure of a strongly correlated system with intermediate dimensionality. Also it is an example compound to study the effect of magnetically active transition metal impurities on the Heisenberg type spin-half antiferromagnetic chains. Magnetic impurities tend to break the long-range magnetic ordering [12]. However, in contrast, there are experimental reports suggesting that when Mn and Fe are doped in CuO, the impurity spins align themselves with the original spin arrangement [13, 14] which needs to be understood.

## 2. Computational details

The density-functional theory (DFT) based first principles electronic structure calculations are carried out using ultra-soft pseudopotentials and plane wave basis sets as implemented in CASTEP simulation package [15] to investigate: (a) origin of non-planarity and its consequence on the stability of ferromagnetic (FM) and antiferromagnetic (AFM) spin-1/2 chains, and (b) the effect impurity spins on these magnetic chains. The exchange-correlation energy is calculated using Perdew-Burke-Ernzerhof general gradient approximation (PWE-GGA) [16]. Additionally Hubbard U is included to account for the strong correlation effect in this transition metal oxide [17]. The value of U is taken as 5 eV unless stated otherwise. To obtain the ground state magnetic ordering, we have performed self-consistent calculations on a $2 \times 1 \times 2$ supercell which consists of 16 formula units of CuO. A $4 \times 2 \times 1$ k-mesh was found to be sufficient for the Brillouin-Zone integration. The plane wave basis set is determined by the cut-off energy of 450 eV. The DFT calculations are also carried out on model structures, as shown in Figure 1, to analyze the cause of non-planarity. For these model structures, appropriate BZ k-mesh are used to achieve the self-consistency.

## 3. Results and discussions

*3.1. Stability and Non-planarity*

Theoretical study on the structural stability of CuO so far is restricted to the metastable tetragonal face-centered structures [18, 19] instead of the ground state monoclinic structure with reduced symmetry [2]. Experimental reports qualitatively attribute the stability of this lowered symmetry structure to severe JT distortion of the cubic rock-salt structure [20]. However, JT distortion in transition metal oxides primarily occurs to remove the band degeneracy and the distortion is of the order of fraction of an angstrom [21]. Therefore, the non-planarity in Cu-Cu, O-O and Cu-O arrangement cannot be explained using JT distortion alone. In this paper, through Figure 1, we have shown the destabilization of the CuO planar structure and the formation of the distorted structure. The experimental stability CuO$_4$ square planar complexes suggest that the crystal structure should naturally be planar as shown in Figure 1(a). However, had the structure been planar, retaining square planar complexes as well as the [101] and [10$\bar{1}$] Cu chains, the oxide would have been a non-magnetic metal as can be seen from the DOS plot of Figure 1(a).

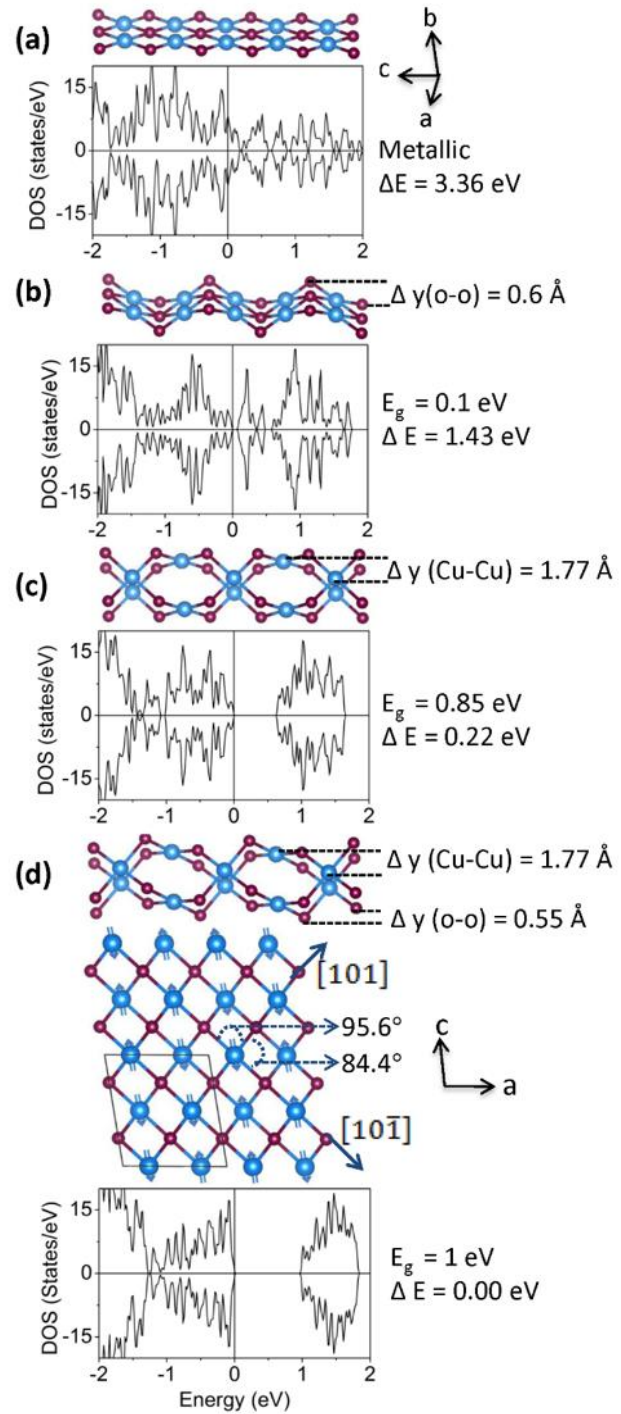

**Figure1.** Hypothetical crystal structures (a, b and c) and the experimental structure (d) of CuO. The corresponding ground state densities of states (DOS) are shown below each structure. For the perfectly two-dimensional square-planar structure (a), a metallic solution is obtained with large DOS at $E_F$. Such a structure increases the electronic kinetic energy which can be significantly reduced by tilting the CuO$_4$ square planar complexes as shown in (b). The structure can be further stabilized by breaking the Cu-Cu planes as shown in (c). The experimental structure (d) is realized by making a minor distortion to the CuO$_4$ complexes of (c). While the planar structure is nearly non-magnetic, rest of the structures are magnetic with experimental spin ordering as shown in (d). Here ΔE represents the relative stability with respect to the experimental AFM ground state of (d). The terms *Δy (O - O)* and *Δy (Cu - Cu)* represent the separation distance between the consecutive O and Cu planes respectively. The Fermi level ($E_F$) is set to zero.

Substantial DOS at the Fermi energy (E$_F$) increases the kinetic energy of the system and thereby it favors lattice distortion, which can also be preferable called as JT distortion, so that the reduced hopping will lead to a gap at E$_F$. Such a configuration can be achieved by displacing the alternate O ions along the b-axis without disturbing the CuO$_4$ plane as shown in Figure 1(b). With displacement ($\Delta y$) of 0.6 Å, a narrow gap opens up at E$_F$ due to reduced O-O covalent interaction. Interestingly, such a distortion not only lowers the total energy of the system by 1.93 eV, but also stabilizes the experimental magnetic ordering which consists of ferromagnetic and antiferromagnetic spin chains along [101] and [10$\bar{1}$] respectively [2, 22]. However, this is not yet the ground state structure. The system prefers to break the Cu-Cu planarity as shown in Figure 1(c) where the Cu and Cu-O layers are stacked alternately along the b-axis. In this way all possible covalent interactions, *viz.* O-O, Cu-Cu and Cu-O-Cu, are significantly reduced to minimize the band energy. As a consequence the total energy of the system is further reduced by 1.11 eV and a wide band gap appears at E$_F$. The experimental ground state structure (Figure 1d) is realized by minor distortion of the CuO$_4$ square planar complex of structure 1(c). We may note that several intermediate structures, not shown here, are examined to understand the evolution of the two-dimensional planar structure to a quasi-three dimensional CuO.

*3.2. Antiferromagnetic Ordering in CuO*

Experimental investigations, via magnetic susceptibility and neutron diffraction measurements, predict AFM ordering in CuO with Néel temperatures 230 and 213 K [22, 23]. Here the 3d$^9$ electronic configuration of Cu$^{2+}$ which can lead to a half-filled band at the Fermi energy. Such half-filled *d*-states undergo Mott-Hubbard transition, due to strong correlation effect, to stabilize the system with an AFM and insulating phase [24, 25]. At the same time, the quasi-one dimensional Cu chains in this compound have inspired many to investigate the spin ordering in CuO using classical Heisenberg model [9, 26] and many other have invoked the phenomenon of Goodenough-Kanamori-Anderson (GKA) rules to explain the AFM ordering in CuO [27]. However, since neither the structure is planar nor the chains are linear, the above theories fail to provide a quantitative estimation of AFM ordering. For example the FM ordering along the [101] chain is contradictory to the spin-half Heisenberg model. In this context DFT calculations are very useful as they have the ability to appropriately evaluate the relation between chemical bonding and magnetic coupling in a compound. The DFT studies on CuO so far have provided varied information on the shape of the magnetic orbital (the orbital that carries the unpaired spin) and therefore the dimensionality of the magnetic coupling. While Filippetti and Fiorentini [8] have attributed $d_{z^2-1}$ as the sole magnetic orbital, Rocquefelte *et al*. [28] disagreed with it and instead reported $d_{x^2-y^2}$ as the magnetic orbital. Therefore, a reexamining of the AFM ordering in CuO using DFT is necessary. This will also form the basis to discuss the impurity spin effect on the pristine magnetic ordering.

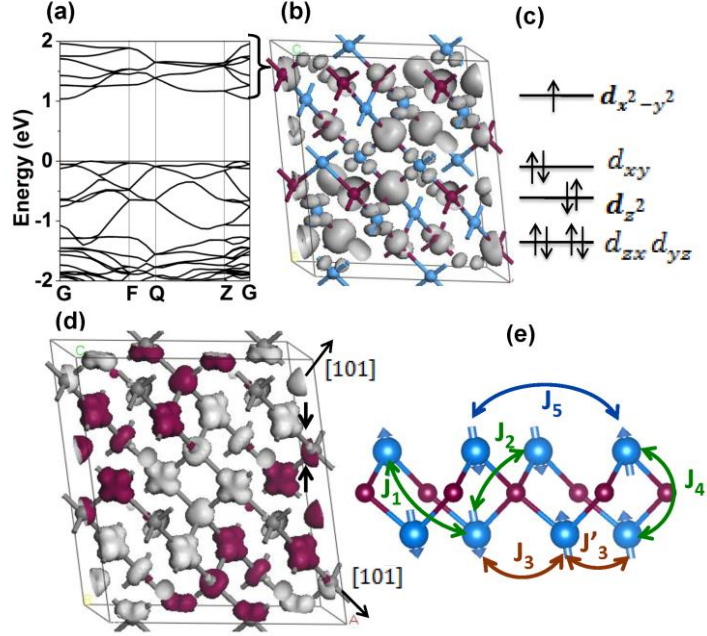

**Figure 2.** (a) Ground state AFM band structure of CuO. (b) Charge density, in the energy range 1 to 2 eV with respect to the Fermi energy ($E_F$), indicating the unoccupied $d_{x^2-y^2}$ orbitals in the spin-minority channel. Here $x$ and $y$ represent the axis of the CuO$_4$ square planar complex. (c) The crystal field split of the $d$ states in a square planar complex. (d) The net spin-density of the system and it is primarily dominated by the spin-majority $d_{x^2-y^2}$ character. Red and grey represent positive and negative spin moments respectively to reflect the AFM [10$\bar{1}$] and FM [101] spin-chains. (e) Various spin-exchange paths estimated in this work.

In Figure 2(a) we have plotted the band structure of AFM CuO. As the considered magnetic unit cell has 16 atoms with 8 in each AFM sublattices, the 3d$^9$ electronic configuration of Cu leads to eight unoccupied spin minority bands lying in the energy range 1 - 2 eV with respect to $E_F$. To identify the character of these bands, we have plotted the corresponding charge density in Figure 2(b). We make two important observations from this figure. (I) Both O-p and Cu-d contribute significantly to the spin density implying reasonable covalent interaction between them. In fact the oxygen magnetic moment is substantial (~ 0.2μ$_B$) in this system. At the same time Cu magnetic moment is close to 0.5 μ$_B$ which is far less than the expected value of 1 μ$_B$. The literature suggests the Cu magnetic moment in the range 0.5 - 0.7 μ$_B$. [22, 29] (II) The shape of the charge cloud concludes that the magnetic orbital is $d_{x^2-y^2}$, where $x$ and $y$ are the local axes of the CuO$_4$ square planar complex. This is in accordance with the crystal field split shown in Figure 2(c). We note that there are two different set of CuO$_4$ complexes in the crystal and each one has different local axes (i.e. axes defining the square planar complex) which also differ from the crystal axes (see Figure 1d). Therefore, the magnetic orbital, $d_{x^2-y^2}$, is a linear combination of the five $d$ orbitals defined with respect to the crystal axes. We shall now present a quantitative measure of the spin-exchange interaction strengths J using Noodelman broken symmetry method [30] as follows. The spin Hamiltonian of a spin dimer with one unpaired spin at each spin site can be written as: $\vec{H} = -J\vec{S}_1 \cdot \vec{S}_2$. Hence, the energy difference between high spin (FM dimer) and low spin (AFM dimer) states is related to J as:

$$E_{hs} - E_{ls} = -J/2 \quad (1)$$

For CuO, we have identified five important exchange interaction paths $J_1$ to $J_5$, shown in Figure 2(e), which determine the magnetic ordering of the system. The $J_i$ are estimated from the simultaneous solution of Eq. (2) which is formulated based on Eq. (1) and DFT obtained total energies of six different magnetic configurations.

$$\begin{aligned}
E_{AF} &= (J_1 - J_2 + 2J_5)/4, \\
E_F &= (-J_1 - J_2 - J_3 - J_4 - 2J_5)/4, \\
E_{AF3} &= (J_1 + J_2 - J_3 + J_4 - 2J_5)/4 \\
E_{AF4} &= (J_1 + J_2 + J_3 - J_4 - 2J_5)/4, \\
E_{AF5} &= (-J_1 + J_2 + 2J_5)/4, \\
E_{AF6} &= (-J_1 - J_2 + J_3 + J_4 - 2J_5)/4.
\end{aligned} \quad (2)$$

Here AF and F respectively represent the experimental AFM ordering as shown in Figure 1(d) and complete FM ordering. In AF3 and AF4, both the chains are antiferromagnetic. In AF3, Cu spins order antiferromagnetically along *c*-axis and ferromagnetically along *a*-axis and reverse is the case for AF4. AF5 consists of FM [$10\bar{1}$] and AFM [101] chains. In the case of AF6, the neighboring spins are antiparallel.

The values $J_i$, are listed in Table 1. The table shows that calculations performed with LDA exchange functional and cluster approximation provide weak magnetic coupling strengths due to inadequate measure of the strong correlation effect. Interestingly, as the table suggests, $J_3$ and $J_4$ can change sign. We find that there are two similar paths for $J_3$ (say $J_3$ and $J_3'$ as in Figure 2(e)). While one of them connects two parallel Cu spins, the other connects two antiparallel Cu spins. Same is the case for the case of $J_4$. These are the consequences of primary exchange paths $J_1$, $J_2$ and $J_5$. The AFM coupling $J_1$ along [$10\bar{1}$] is the strongest and it resembles to that of a one-dimensional spin-1/2 Heisenberg AFM chain. Contrary to the similar expectation, the coupling $J_2$ along [101] is weak and ferromagnetic. Earlier, this contradiction is qualitatively explained by invoking the GKA rules [33-35]. According to this rule, the coupling is antiferromagnetic if Cu-O-Cu bond angle is 180° and ferromagnetic if the bond angle is 90°. However, as shown in table-I, none of the bond angles are either 180° or 90°. For $J_1$, the angle is 139° and for $J_2$, it is 106°. Therefore, an alternative explanation is required. The charge density plotted in Figure 2b, shows that the p-d covalent interaction is almost absent along [$10\bar{1}$]. As the d-d interaction is expected to be negligible by virtue of being a second neighbor interaction, the spin Hamiltonian is completely Heisenberg type leading to an AFM spin chain along [$10\bar{1}$]. On the contrary, there is a weak covalent interaction between the $d_{x^2-y^2}$ and $p_x$ and $p_y$ states along [101] which is sufficient enough to overcome the AFM coupling and to favor a kinetic exchange driven weak FM ordering. The large value of $J_5$ (-39.11 meV) reflects the three-dimensional nature of the magnetic unit cell.

**Table 1.** Exchange interaction parameters ($J_i$ in meV) estimated from Eq. 2. Some of the literature data are also listed here for comparison. We note that as per experimental studies, the net AFM interaction strength is -77 meV [22, 23]**.** The Cu-Cu bond length (d) and Cu-O-Cu bond angle (θ) for each path are also indicated.

| [Reference] Methodology | $J_1$ d = 3.99 Å θ = 139° | $J_2$ d = 3.89 Å θ = 106° | $J_3$ d = 3.95 Å θ = 95° | $J_4$ d = 3.93 Å θ = 106° | $J_5$ d = 7.86 Å -- |
|---|---|---|---|---|---|
| Present Work Pseudopotential (GGA+U) | -127.48 | 8.6 | -33.18 | -3.29 | -39.11 |
| FP-LAPW (Hybrid/GGA) [28] | -128.8 | 2.6 | -18.2 | 4.2 | -30.1 |
| PAW (GGA+U) [31] | -107.7 | 15.76 | -15.82 | -7.98 | -16.18 |
| Pseudopotential (LDA-SIC) [11] | -38.4 | 20.4 | 8 | 11.6 | -14 |
| Cluster-SCF (active space) [32] | -10 | 2.9 | 7.5 | 3.4 | -- |

*3.3. Impurity Spin Effect*

In general perturbation due to impurities and defects tends to create local spin disorder by breaking the long range magnetic ordering. However, experimentally it is observed that dilute Mn doping (less than 0.2 %) in CuO does not break the AFM chains [36] which is also supported by DFT studies [11]. Similarly in the case of Fe doping, theoretically it is shown that Fe inherits the Cu spin and makes the system ferrimagnetic [13]. With Ni doping, experimentally it is found that while the Néel temperature is not affected [37], there might be the possibility of short range ordering in the Ni doped region. The cause of sustaining long-range order in Mn and Fe doped CuO and probable short-range order in Ni doped CuO has not been explained yet. In the present study we have provided a mechanism explaining the effect of transition metal (Cr, Mn, Fe, Co and Ni) impurity on the Cu-spin chains in this 2D-3D crossover system.

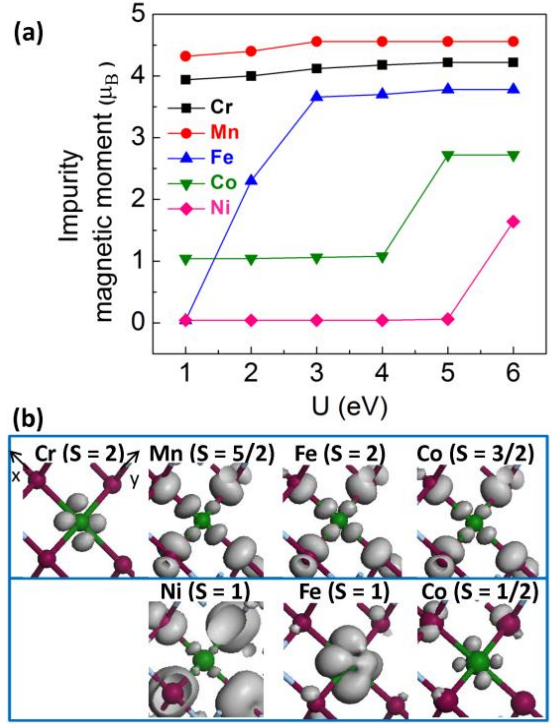

**Figure 3.** (a) Local magnetic moment at the impurity site (M) as a function of U. Mn and Cr stabilize only with HS configuration. Ni and Co make a transition from LS to HS at higher value of U. Iron can exhibit all of the spin states. (b) The impurity charge density in the vicinity of $E_F$ (1 eV) for LS, IS and HS states of the impurities. The LS state of Fe and Ni are magnetically inert and the occupied d-states are lying far below $E_F$. Hence they are not shown in the figure.

By varying the dopant from Cr to Ni, we can examine the effect of $3d^4$ to $3d^8$ impurities on the spin-1/2 chains of CuO. In Figure 3a, we have shown the impurity spin states of the dopants as a function of Hubbard U. With U, the increase in localization makes the d states more spin-polarized which in turn increases the magnetization. In the case of Cr and Mn, we always have a high spin state with S = 2 and 5/2 respectively. However, the orbital occupancy differs for these two dopants. While $d_{x^2-y^2}$, the uppermost orbital in the energy level due to crystal field split, is occupied in the majority spin channel in Mn, it is completely empty in Cr (see Figure 2c and 3b). The ground state, obtained from the DFT calculations, shows that the Mn impurity spin retains the spin orientation of the replaced Cu which is in agreement with the literature [14]. However, the Cr impurity spin reverses the spin-orientation and hence breaks the spin chains. This significantly implies that the occupancy of impurity $d_{x^2-y^2}$ orbital plays a major role in affecting the long-range magnetic ordering of the host compound. To substantiate it further, we have studied the spin state of other impurities as well. According to Figure 3a, three possible spin states, *viz.*, low spin (LS; S = 0), intermediate spin (IS; S = 1) and high spin (HS; S = 2), can be realized for $Fe^{2+}$ ($3d^6$) by varying the value of U. While for HS, the magnetic orbital is $d_{x^2-y^2}$, it is $d_{xy}$ for the IS configuration. We find that HS Fe inherits the spin orientation of the replaced Cu and IS Fe prefers to flip the spin at the impurity site. For LS state, Fe spin moment is zero and therefore the impurity can break the long-range magnetic ordering to introduce short range magnetic domains [38]. The LS and HS states of the Co impurity ($3d^7$) are realized with S = 1/2 and 3/2 respectively. As per the crystal field split (see Figure 2c), the LS implies the magnetic orbital as $d_{xy}$ which is also reflected in our spin density plot in Figure 3b. This leads to a spin flip at the impurity site as in case of IS Fe. However, the HS configuration makes the $d_{x^2-y^2}$ orbital half-occupied and it retains the parent spin alignment. Similarly with Ni impurity, the HS configuration favors the host magnetic ordering with $d_{x^2-y^2}$ being the uppermost spin-polarized orbital. On the other hand, the Ni-LS is magnetically inactive like Fe-LS.

While discussing the magnetic ordering of CuO, we have shown that if the magnetic orbital, i.e. the top most spin-polarized state, is $d_{x^2-y^2}$ then the weak but robust covalent interaction between this orbital and O-p orbitals along [101] leads to FM ordering. The absence of such an interaction along [10$\bar{1}$] makes the chain antiferromagnet. The same is observed for the impurity spin alignment. For demonstration we have projected the spin density of $Cu_{0.9375}M_{0.0625}O$ for low and high spin configurations of impurity M in Figure 4. It shows that, in the HS state, Mn, Fe, Co and Ni impurities retain the replaced Cu spin orientation. According to Figure 3(b), in all of these cases the magnetic orbital is $d_{x^2-y^2}$. The HS Cr, IS Fe and LS Co have opposite spin orientation with respect to the replaced spin. The magnetic orbitals of these configurations differ from $d_{x^2-y^2}$.

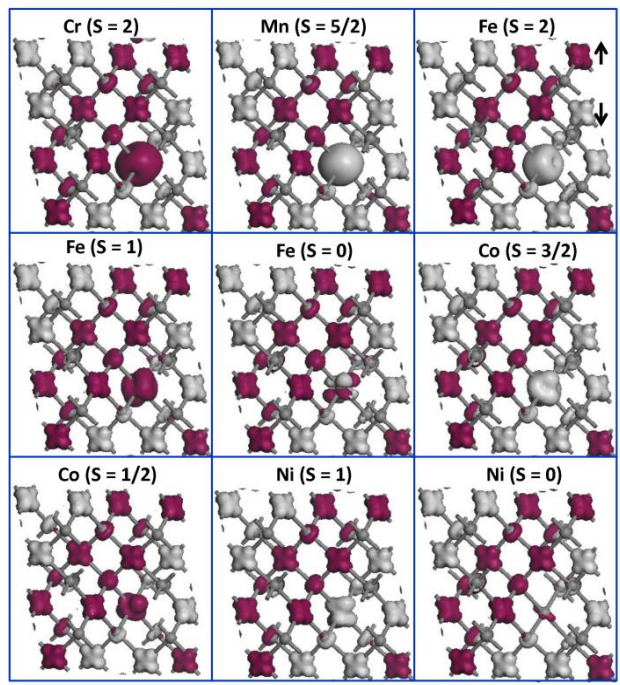

**Figure 4.** Spin density for $Cu_{0.9375}M_{0.0625}O$ (M = Cr, Mn, Fe, Co and Ni). The grey and maroon represent opposite spin as indicated. The alignment of the impurity spin and its position can be easily seen from the figure. The results are obtained for single impurity in a magnetic unit cell consisting of 16 formula units.

When the magnetic orbital differs from $d_{x^2-y^2}$, the absence of covalent interaction tends to make Cu-M interaction antiferromagnetic both along $[10\bar{1}]$ and $[101]$. However, both being antiferromagnetic, it leads to spin-frustration as can be seen from Figure 1d. The magnetic coupling strength will be stronger along $[101]$ than along $[10\bar{1}]$ as the former has smaller Cu-M bond length (~3.89 Å) than the latter (4Å). Therefore, the AFM interaction survives along $[101]$ forcing a spin-flip at the impurity site. The strength of magnetic interaction between host and impurity spins can be measured from the energy difference $\Delta E$ (= $E_{NFL} - E_{FL}$) between the magnetic configuration with impurity retaining the pristine spin alignment and the state with flipped impurity spin. The value of $\Delta E$ for different possible impurity spin states is listed in Table-II. The table further substantiates the role of the magnetic orbital in host-impurity interaction in CuO.

**Table 2**. Energy difference $\Delta E = E_{NFL} - E_{FL}$, where $E_{NFL}$ and $E_{FL}$ are the total energies of $Cu_{0.9375}M_{0.0625}O$ with impurity M inherits and the flips the replaced Cu spin respectively, for all possible impurity spin states (see Figure 3). The value of $\Delta E$ for the pure compound (i. e. M = Cu) is -206 meV.

| $\Delta E$ (meV) | Cr | Mn | Fe | Co | Ni |
|---|---|---|---|---|---|
| | 116.4 (HS) | -151.8 (HS) | -174.5 (HS) | -190 (HS) | -132.2 (HS) |
| | | | 75.1 (IS) | 43 (LS) | |

## 4. Summary

To summarize, we have examined the structural stability of CuO and 3d impurity spin effect on the magnetic chains of the host compound. The monoxide undergoes severe lattice distortion in order to minimize the *p-p* and *p-d* and d-d covalent interactions. The reminiscent $p - d_{x^2-y^2}$ covalent interaction is sufficient enough to make the spin chain along $[101]$ ferromagnetic while the other along $[10\bar{1}]$ remains Heisenberg type antiferromagnet. The magnetic alignment of the impurity spin with the host spin is purely orbital driven. If the top most occupied impurity spin orbital is $d_{x^2-y^2}$, the original antiferromagnetic ordering is maintained. Else the flipped impurity spin breaks the long range ordering. The host-impurity interaction mechanism evolved from this work is unique with respect to the same in three dimensional and low dimensional magnetic systems and can be a prototype to explain the impurity magnetic phenomena in other systems stabilizing with intermediate dimensionality.

**Acknowledgement**

We acknowledge Department of Biotechnology (DBT), India for the financial support and HPCE of IIT Madras for the computational resources.